\documentclass[
hf
]{ceurart}

\usepackage[normalem]{ulem}

\begin{document}

\copyrightyear{2020}
\copyrightclause{Copyright for this paper by its authors.\\
  Use permitted under Creative Commons License Attribution 4.0
  International (CC BY 4.0).}

\conference{ROMCIR 2021: Workshop on Reducing Online Misinformation through Credible Information Retrieval, held as part of ECIR 2021: the 43rd European Conference on Information Retrieval, March 28 – April 1, 2021, Lucca, Italy (Online Event)}

\title{Supporting verification of news articles with automated search for semantically similar articles}



\author[1]{Vishwani Gupta}
\ead{vishwani.gupta@iais.fraunhofer.de}

\author[1,2]{Katharina Beckh}
\ead{katharina.beckh@iais.fraunhofer.de}

\author[1,2]{Sven Giesselbach}
\ead{sven.giesselbach@iais.fraunhofer.de}
\ead[url]{http://www.iais.fraunhofer.de/nlu}

\author[1]{Dennis Wegener}
\ead{dennis.wegener@iais.fraunhofer.de}
\ead[url]{http://www.iais.fraunhofer.de/mlops}

\author[1,3]{Tim Wirtz}
\ead{tim.wirtz@iais.fraunhofer.de}

\address[1]{Fraunhofer IAIS}
\address[2]{Competence Center for Machine Learning Rhine-Ruhr}
\address[3]{Fraunhofer Center for Machine Learnning}


\begin{abstract}
Fake information poses one of the major threats for society in the 21st century. 
Identifying misinformation has become a key challenge due to the amount of fake news that is published daily. Yet, no approach is established that addresses the dynamics and versatility of \emph{fake news editorials}. Instead of classifying content, we propose an evidence retrieval approach to handle fake news. The learning task is formulated as an unsupervised machine learning problem. For validation purpose, we provide the user with a set of news articles from \emph{reliable news sources} supporting the hypothesis of the news article in query and the final decision is left to the user. 
Technically we propose a two-step process: (i) \emph{Aggregation}-step: With information extracted from the given text we query for similar content from reliable news sources. (ii) \emph{Refining}-step: We narrow the supporting evidence down by measuring the semantic distance of the text with the collection from step (i). The distance is calculated based on Word2Vec and the Word Mover's Distance. In our experiments, only content that is below a certain distance threshold is considered as supporting evidence. We find that our approach is agnostic to concept drifts, i.e. the machine learning task is independent of the hypotheses 
in a text. This makes it highly adaptable in times where fake news is as diverse as classical news is. Our pipeline offers the possibility for further analysis in the future, such as investigating bias and differences in news reporting. 
\end{abstract}


\maketitle

\section{Introduction}
Although its negative influence and its weaponizing usage is known for ages \cite{soll2016long}, \emph{fake news} (in non-political context also known as \emph{false news}~\cite{wardle2017information}) and its negative impact was globally recognized, during the U.S. elections in 2016, as one of the major challenges for the society of the 21st century \cite{activities2017intentions,howell2013digital}. Importantly, it was used to promote both political campaigns in the election. Beside the promotion of political campaigns~\cite{barstow2008behind,allcott2017social}, fake news occurs with various purposes or due to various circumstances, e.g. to destabilize governments in third countries, accidentally due to unconscious misinterpretation of facts~\cite{quandt2019fake} or as a worthwhile revenue stream based on advertisement~\cite{waldrop2017news,kshetri2017economics,braun2019fake}.

Following~\cite{quandt2019fake} the term \emph{fake news} is used twice (a) to discredit and downgrade media and journalism; and (b) to summarize various forms of wrong, misguided, or fabricated information. Throughout this manuscript we are speaking about (b) when discussing fake news. Fake news articles, as just described, are to a large extend published, maintained, circulated and promoted in social media~\cite{wingfield2016google,koohikamali2017information,bovet2019influence,shu2019studying}. 
On a high level, two strategies of potential interventions have been highlighted~\cite{lazer2018science}, (i) empowering of individuals to evaluate and assess fake news and (ii) structural changes preventing exposure of fake news to individuals. Most likely, machine learning based intervention strategies can be categorized into the second class of strategies. However, our intent is to propose an approach, while mainly based on state-of-the-art machine learning methodology, that can be categorized into the first class. Because we believe that the most sustainable strategy to fight the impact of fake news is to empower individuals to evaluate and assess fake news, we propose to assess content with evidence from reliable news sources supporting the hypotheses in the articles. We leave the final decision to the user which helps to improve acceptance because no actual censorship is happening. However, a quantitative statistical evaluation is still possible by simply adding a threshold from cross-validation experiments on top of the mechanism. 

In summary our contribution can be structured into the following aspects:
\begin{itemize}
    \item Modular system for the comparison of news articles from various sources.
    \item Unsupervised approach for verification of a queried article and its content. 

    \item Automatic querying for supporting articles using News API.
    \item An intuitive user interface which allows to  individualize the collection of reliable sources and to receive visual feedback for the queried article.
\end{itemize}

The outline of the paper is as follows: Section \ref{sec:related_work} summarizes the main related work, highlighting prior approaches towards verification. In section \ref{sec:building_blocks}, we outline the relevant machine learning building blocks of our approach. Section \ref{sec:appoach_details} introduces our system for news verification and describes the workflow, architecture and user interface. The deployment of the solution architecture is described in section \ref{sec:architecture} as well as a discussion of the approach in general and its advantages in section \ref{sec:discussion}. Finally, in section \ref{sec:conclusion} we summarize the approach.

\section{Related Work}\label{sec:related_work}

Following the line of argumentation of \cite{zhou2020survey}, approaches to identify fake news can be structured into four major categories: knowledge-based, style-based, propagation-based and source-based 
Propagation-based analyses are concerned with how fake news spread online which is mostly formulated as a binary classification problem. The input can be either a news cascade \cite{castillo2011information} or a self-defined graph \cite{jin2016news}.
For style-based analysis, the writing style is assessed according to malicious intent. Perez et al.~\cite{umich} point out stylistic biases that exists in text in order to automate fake news detection. Source-based approaches assess the credibility of a news source \cite{newsguard, baly2018predicting} while knowledge-based approaches compare news content with known facts \cite{botnevik2020brenda}.

According to the scheme from Zhou et al.~\cite{zhou2020survey}, our proposed approach combines two categories of fake news detection: source-based and knowledge-based analysis. For both, we highlight prior work.

The most prevalent source-based approach is to rate news sources on their credibility. Traditional source-based approaches are Web ranking algorithms which rely on website credibility to improve search results for user queries \cite{page1999pagerank}. Two current resources for news publisher credibility are MediaBias/FactCheck \cite{mediabiasfactcheck} and NewsGuard~\cite{newsguard}, a browser extension that displays ratings of news websites. The ratings are manually curated by journalists. Since the ratings go through a manual review process the list of rated websites is prone to be incomplete and quickly outdated. Therefore, recent efforts aim for automating source reliability ratings. Based on \emph{expert-}features including for example web-traffic, the existence of a verified Twitter account or textual information, the authors of~\cite{baly2018predicting} classify the news sources in a supervised manner using a Support Vector Machine. 
Another approach to evaluate credibility of the knowledge is proposed by Esteves et al. \cite{jens}. The proposed approaches are based on supervised learning to automatically extract source reputation cues and to compute a credibility factor.
A further approach that also taps into style-based methods is to analyse text and metadata in the article. Rashkin et al. \cite{rashkin} assess the reliability of entire news articles by predicting whether the document originates from a website classified as hoax, satire or propaganda by comparing the language of real news with those three categories to find linguistic characteristics of untrustworthy text. Wang et al. showed that significant improvements can be achieved for fine-grained fake news detection when meta-data is combined with text \cite{wang}.

Knowledge-based approaches mostly tackle the process of fact-checking. Several fact-check-ing organizations such as CORRECTIV~\cite{correctiv}, PolitiFact~\cite{politifact} and Snopes~\cite{snopes} operate by manually verifying claims (see \cite{zhou2020survey} for more expert-based fact-checking websites). A drawback of manual verification is that it may reach readers too late. An approach tackling this issue was recently published~\cite{botnevik2020brenda} where the authors approached fact-checking with machine learning methods and focuse on claim verification. 

Another related approach is presented in \cite{fever} where the authors introduce a Fact Extraction and VERification (FEVER) Shared Task. The aim is to classify whether a claim is factual or not by retrieving evidence from Wikipedia.
Both works treat the task as a classification problem, and a critical challenge with this approach is that we can not guarantee that the system is able to give suggestions to very recent claims. 
An approach geared towards misinformation detection for social media treating exactly this challenge is presented in~\cite{hossain2020covid} by including a retrieval step.
While the authors still include classification as a second step, i.e. for stance detection, we completely omit any supervised task and focus on retrieval and an expert-knowledge-based scoring. 

\section{Building Blocks of the Approach}
\label{sec:building_blocks}
As presented in more detail in section \ref{sec:appoach_details} we propose an evidence retrieval approach to handle fake news instead of classifying content. The learning task is formulated as an unsupervised machine learning problem. The evidence supporting the hypothesis of the queried article is gathered from a collection of reliable news sources which we provide to the user. It is individualized by  selecting an arbitrary number of sources out of a curated list of reliable news sources for evidence-gathering purposes.
Technically, we propose a two-step process:
\begin{enumerate}
    \item \emph{Aggregation}-step: Extract information from the given article and query for similar content from reliable sources 
    \item \emph{Refining}-step: Narrow the supporting evidence down by calculating the semantic distance of the text with the collection that was retrieved in step 1.
\end{enumerate}

In the following subsection we briefly introduce the most relevant machine learning concepts, forming the basis of the proposed approach. To calculate the semantic distance of news articles we rely on distributed word embeddings and the Word Mover's distance.

\subsection{Word Embedding}
Mikolov et al. \cite{mikolov2013efficient} proposed the Word2Vec algorithm to learn vector representations of words. The method is based on the distributional hypothesis \cite{harris54,Firth_Papers57} that words get their meaning from the context in which they appear. Mikolov et al. propose two different variations of the Word2Vec algorithm, both typically trained on large text corpora. The Continuous Bag-of-Words Model (CBOW) and the Continuous Skip-gram Model (skip-gram) which predict target words from source context words and source context words from target words, respectively. Specifically, they propose a shallow neural network architecture, which trains continuous word vectors representations  to maximize the log probability of neighboring words in a corpus. For a given sequence of words $w_1, w_2, ..., w_N$, it models the probability of this particular sequence as follows 
\begin{equation}
    \frac{1}{N}\sum_{n=1}^{N}\sum_{j\in nb(n)} \log p(w_j\mid w_n)
\end{equation}
Here, $nb(n)$ is the set of neighboring words of the word $w_n$. The unsupervised training is done by optimizing the maximum likelihood of a corpus of sentences (sequences of words) such that the word embeddings capture the semantic information of words and relations between them, given a particular context. In their original work~\cite{mikolov2013efficient}, the authors approximated the objective above by more efficiently trainable objectives. 

A flaw of Word2Vec is its inability to infer continuous representations for words not seen during training. Especially in domains such as news, new vocabulary can emerge rapidly. A simple way to account for that is to incorporate morphological information about words in the text representations. Bojanowski et al. \cite{bojanowski2016enriching} proposed \emph{fastText}, an extension of the skip-gram model, which learns word representations by including sub-word information. This is achieved by not only representing words with vectors but also the sub-word parts they consist of, bag of character \textit{n}-grams. Word vector representations are built as the sum of their sub-word and their own representation.

In this work, we experimented with two embedding models, Word2Vec and fastText embeddings.

Although there are by far more than two approaches available in the literature (also more advanced approaches like Transformers~\cite{vaswani2017attention}), see \cite{li2018word,almeida2019word} for comprehensive reviews, we focus on those because they can be efficiently implemented on standard hardware and are well-established in the NLP community.  
Nevertheless, there is freedom in experimenting with other word embeddings as it only requires a change of the distance threshold. Hence, the approach can be easily adapted to support news verification for different languages. 

\subsection{Word Mover's Distance}
Earth mover's distance (EMD), also known as the Wasserstein distance, is a distance measure between two probability distribution. Kusner et al. \cite{kusner2015word} proposed a version of EMD applicable to language models, the Word mover's distance (WMD) which evaluates the distance between two documents represented in a continuous space using word embeddings such as the aforementioned Word2Vec and fastText embeddings. For any two documents \textit{A} and \textit{B}, WMD is defined as the minimum cost of transforming document \textit{A} into document \textit{B}. Each document is represented by the relative frequencies of its words relative to the total number of words of the document, i.e., for the \textit{j}th word in the document, 
 \begin{equation}
    d_{A,j} = count(j)/\mid A \mid
    \end{equation}
where $\mid A \mid$ is the total word count of document A and $count(j)$ is number of occurrences of the word with vocabulary index $j$. The \textit{j}th word is represented by its corresponding word embedding, say $\mathbf{v}_j \in \mathbb{R}^{n}$. The $n$-dimensional word embeddings are obtained from a pre-trained model, e.g. Word2Vec or fastText. The distance between two words can easily be measured using Euclidean distance,
\begin{equation}
    \delta(i,j) =\lVert \mathbf{v}_i - \mathbf{v}_j \rVert
\end{equation}
Based on this choice, the Word mover's distance is defined to be the solution of the following linear program,
\begin{equation}
\begin{aligned}
    WMD (A,B) =\quad & \min_{\mathbf{T}\geq0}\sum_{i=1}^{V}\sum_{j=1}^{V}\mathbf{T}_{i,j} \delta(i,j)\\
    \textrm{such that}\quad & \sum_{i=1}^{V}\mathbf{T}_{i,j}=d_{A,j} \\
    \textrm{and}\quad & \sum_{j=1}^{V}\mathbf{T}_{i,j}=d_{A,i} \\
\end{aligned}
\end{equation}
Here, $\mathbf{T} \in \mathbb{R}^{V \times V}$ is a non-negative matrix, where $\textbf{T}_{i,j}$ denotes how much of word \textit{i} in document \textit{A} is assigned to tokens of word \textit{j} in document \textit{B}. 
Empirically, WMD has reported improved performance on many real world classification tasks as demonstrated in \cite{kusner2015word}. The WMD has intriguing properties. The distance between two documents can be broken down and represented as the sparse distances between few individual words. The distance metric is also hyper-parameter free. The most important feature is that it incorporates the semantic information encoded in the word embedding space and is agnostic to arbitrary word embedding models.

\section{A Retrieval-Based Approach Supporting Fake News Identification Methods}\label{sec:appoach_details}

\begin{figure}
    \centering
    \includegraphics[scale=0.78]{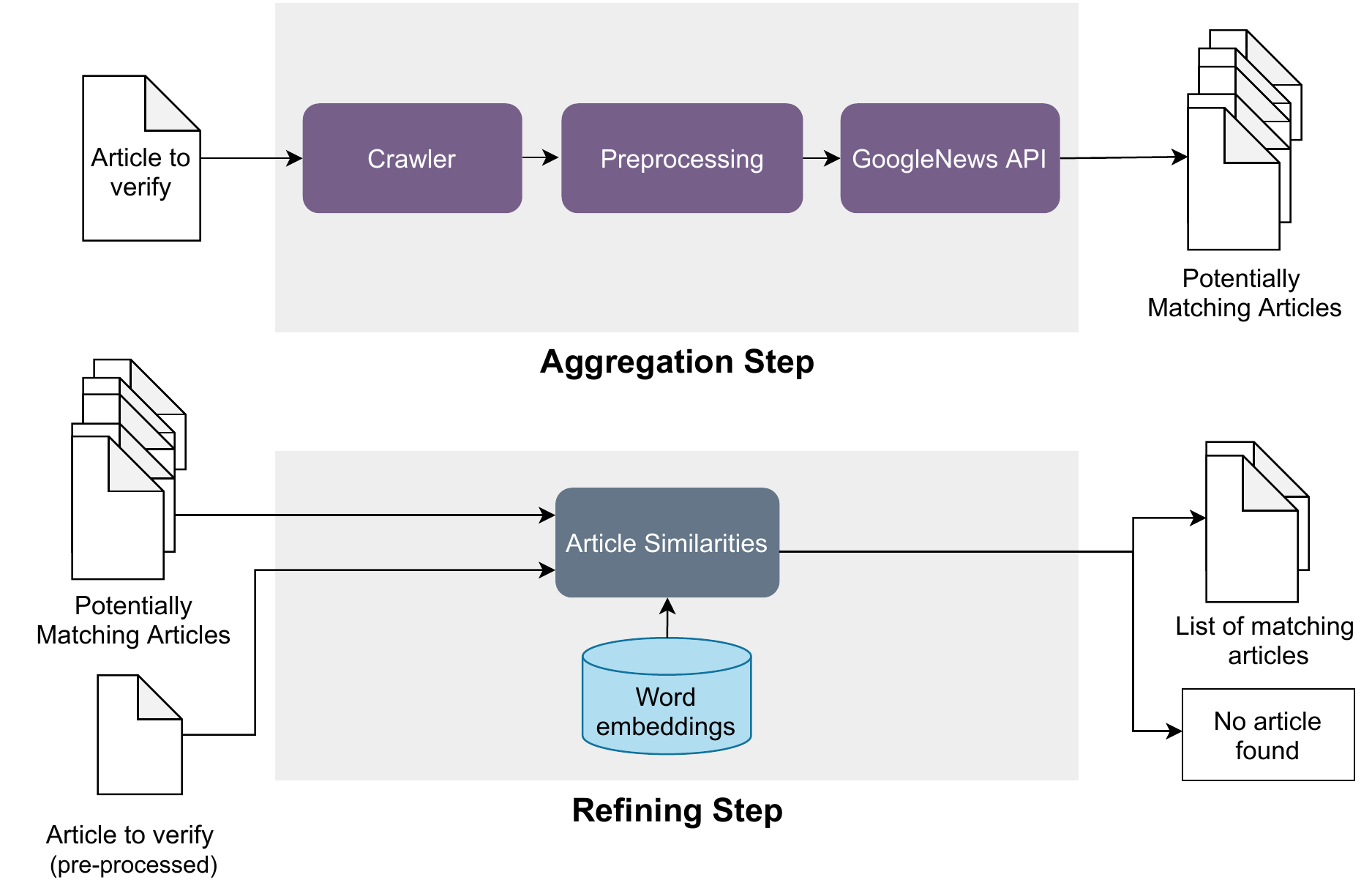}
    \caption{Overview of the system's workflow with aggregation and refining step.}
    \label{fig:system}
\end{figure}

We constructed a pipelined system which helps in extracting semantically similar articles from reliable news sources. Its core is the analysis of the credibility of news articles based on the overall evidence collected from a set of automatically retrieved articles published by reliable news sources. 
Figure \ref{fig:system} gives an overview of the components and workflow. The system consists of three components: a news content extractor, a search engine query and a content analyzer. All of these components can easily be exchanged and extended depending on the language and the list of reliable news sources.

\subsection{Article Extractor}
Given a link to an article that should be verified the article extractor component extracts information from the link such as the publication date, the article title, its authors and the textual content. For most of the news sources, the python library Newspaper3k\footnote{https://newspaper.readthedocs.io/en/latest/} is suitable and manages to extract all of the above information. However, at least for some sources, we built our own article extractors by parsing the HTML page of the article and extracting certain tags. 
We extract relevant keywords and entities from the title and body of the article using the Newspaper API keyword extractor. The goal of this step is to get as much relevant information characterizing the article as possible. This extracted information is used in building an automated query.

\subsection{Querying the Google News API}
To obtain news articles from reliable news sources we use a query component which queries the Google News API. Using the keywords and entities we obtained in the step before, we construct a query. We structure the query so that we can filter the articles based on date, number of requested news articles from the sources, location and language. The API returns ten article links based on the search criterion from every source selected. The number of articles returned by the API can be changed based on individual requirements and computation power. The article extractor component extracts the content of the articles obtained from the search API. 
The system offers six news sources and we can easily add new sources or remove existing ones from the list. Automatic querying used here is different from manual news search using search engines as we aggregate news based on dates, keywords extracted from the article, and selected reliable sources.

\subsection{Content Analysis: Semantic Distance Analysis}
The content analysis component computes the semantic distance between the query article and the articles returned by the query component. Before computing the distance score, we clean the article titles and bodies by removing stop words and special symbols and computing their bag of n-grams representations. The semantic distance of articles is calculated using word embeddings and the WMD. For the word embeddings, we experimented with different word embeddings such as fastText and the pre-trained Google news embeddings. The quality of the word embeddings depends on the size of training data, thus, we use pre-trained word embeddings.

Since the original WMD is computationally expensive, we approximate the distance by using the Regularized Wasserstein distance proposed by \cite{regWMD} and only keep the five closest articles. The five articles with the least distance are then selected for computation with the original WMD. The WMD returns a distance score for each remaining article from the individual sources. The smaller the distance, the more related the articles are. Only articles that are below a predefined threshold are considered as similar to the given article. We set the distance threshold by empirically checking the distances of a couple of articles. Similar news articles, i.e. articles that fall below the distance threshold, are then displayed with a message that closely related articles were found. If the system does not return similar articles the reader is informed that the given article is potentially fake. 

Our prototype was exemplary tested on a small set of articles. A systematic evaluation with a self-curated dataset and the FakeNewsNet \cite{data} dataset is planned. 
The semantic distance analysis in our approach is based on unsupervised models which in turn make the system highly adaptable to different languages. We just need to replace the word embeddings and adapt the threshold. 

Furthermore, the unsupervised nature renders the approach agnostic to concept drifts which means that the machine learning task is independent of the hypotheses in a text.

\section{Architecture and Deployment}\label{sec:architecture}

\begin{figure*}
    \centering
    \includegraphics[scale=0.9]{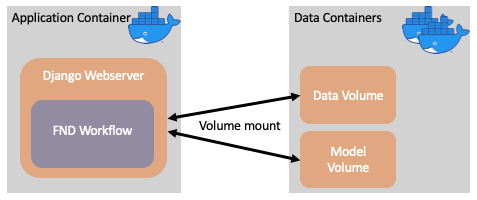}
    \caption{The technical architecture of the Fake News Detector with three docker components.}
    \label{fig:architecture}
\end{figure*}

To showcase our approach we build a "Fake News Detector" system. The Fake News Detector system 
consists of a few technical components. Its technical architecture is based on a set of docker components (see Figure \ref{fig:architecture}). In detail, there are the following three docker components:
\begin{enumerate}
    \item A container with simple django running the python code of our application and serving the frontend.
    \item A container serving the data for the backend - the model container.
    \item A container that includes data pre-processed by several NLTK functions.
\end{enumerate}

\begin{figure*}
    \centering
    \includegraphics[width=\textwidth]{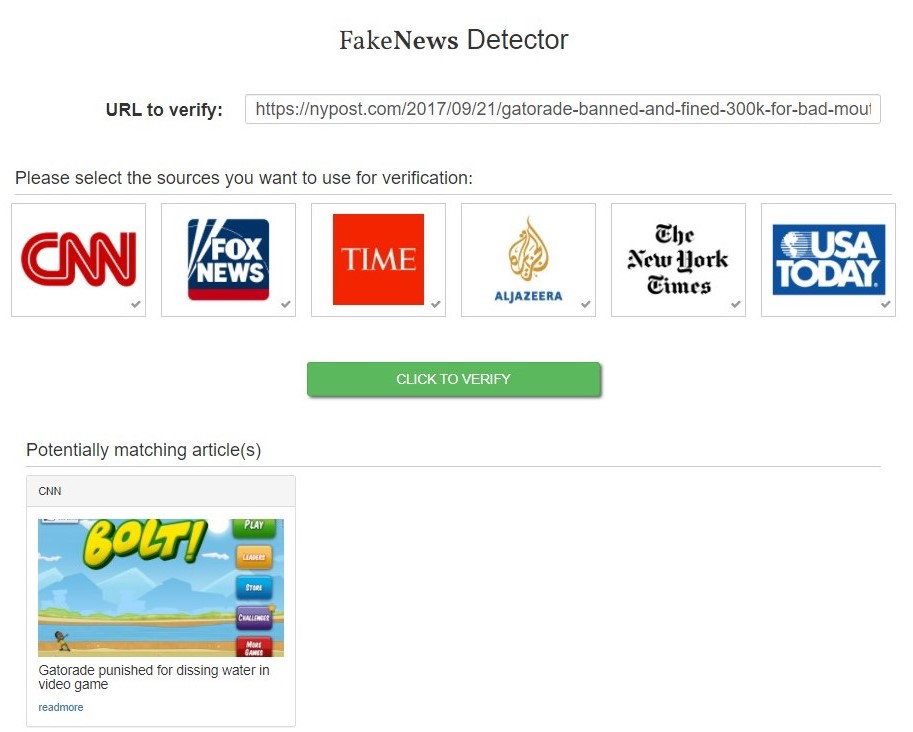}
    \caption{The User Interface of the Fake News Detector. In this example, the query article is titled \textit{Gatorade banned and fined \$300k for bad-mouthing water}. The user pastes the URL into the text box, selects sources and receives a matching article from the source CNN. }
    \label{fig:ui}
\end{figure*}

The Fake News Detector application can be accessed via web UI (see Figure \ref{fig:ui}). In the UI, we can insert a link of a news article to be verified, in this example, we want to verify an article titled \textit{Gatorade banned and fined \$300k for bad-mouthing water}. Next, we select the news sources to check and match against. By clicking on the verification button the analysis process is started at the backend running all the components. After the analysis, the results are shown as a list of potentially matching articles. If no matching articles are found after the analysis, a message is displayed that the article might be potentially fake. In the example shown in Figure \ref{fig:ui}, we selected all six sources. The system queried against all sources and analysed the potentially matching articles using semantic similarity and found that CNN has published a similar news article during the same time frame.

\section{Discussion and Future Work}\label{sec:discussion}

In the previous sections, we have addressed various benefits of following such an unsupervised approach. These benefits are the plasticity of the system, its modularity and user-driven decision making. In this section, we discuss several challenges and insights for the future work.

One challenge for the demonstrator is to deal with very recent news. The system will not be able to collect semantically similar articles from other reliable news sources that might not have published yet on the subject. Here, date as well as publishing time become important. For future work, we propose to highlight this fact to the user to avoid perceiving recent news as potentially fake.

Another challenge in analyzing news articles is novel words which are out-of-vocabulary. For these words no word vectors exist which in turn affects the WMD score. The system can handle this scenario but the performance may be impaired. To overcome this we are crawling news from major news sources every hour to build a text dataset. We propose to update word embedding models frequently to avoid an increased number of out-of-vocabulary words in recent news articles.

Recently, Yokoi et al. \cite{yokoi2020word}, proposed an improvement on Word Mover's Distance, namely Word Rotator's Distance (WRD) which measures the degree of semantic overlap between two texts using word alignment. This approach is designed such that the norm (a proxy for the importance of word) and angle of word vectors (a proxy for dissimilarity between words) correspond to the probability mass and transportation cost in EMD, respectively. This approach outperforms the WMD in several semantic textual similarity tasks. This alternative might help in improving the document distance threshold and is subject for our planned evaluation.

We also propose to use the approach of document similarity for related use cases where we see potential in two directions. 
The first direction is helping fact-checkers by providing adequate evidence to verify hypotheses. By providing them similar content, e.g. evidence in the form of news but also scientific articles, the system can support their task. 
Second, reviewers in several domains, e.g. medical health news review, need to determine how comprehensible a text document is. By comparing a text to scientific or more simple language it is possible to provide a comprehensibility score. 

Since the system alone does not guarantee news verification or falsification we recommend considering combining it with fact-checking methods.

\section{Conclusion}\label{sec:conclusion}
We presented a system to find semantically similar articles to a given news article from selected reliable sources. For the system, we propose an evidence retrieval approach to handle fake news instead of treating it as a classification task. This way, we aid the users in finding supporting evidence and, thus, manual search work can be reduced. 

The benefits of our system are that (i) it is unsupervised and therefore agnostic to concept drifts, (ii) it gives the user decision power and  (iii) it is modular, i.e. the system can be easily adapted to other languages, extended and improved with further components.

\begin{acknowledgments}
This research has been partly funded by the Federal Ministry of Education and Research of Germany as part of the Competence Center for Machine Learning ML2R (01|S18038B).
T. Wirtz  contributed as part of the Fraunhofer Center for Machine Learning within the Fraunhofer Cluster for Cognitive Internet Technologies.
\end{acknowledgments}

\bibliography{ref}

\begin{thebibliography}{43}
\expandafter\ifx\csname natexlab\endcsname\relax\def\natexlab#1{#1}\fi
\providecommand{\url}[1]{\texttt{#1}}
\providecommand{\href}[2]{#2}
\providecommand{\path}[1]{#1}
\providecommand{\DOIprefix}{doi:}
\providecommand{\ArXivprefix}{arXiv:}
\providecommand{\URLprefix}{URL: }
\providecommand{\Pubmedprefix}{pmid:}
\providecommand{\doi}[1]{\href{http://dx.doi.org/#1}{\path{#1}}}
\providecommand{\Pubmed}[1]{\href{pmid:#1}{\path{#1}}}
\providecommand{\bibinfo}[2]{#2}
\ifx\xfnm\relax \def\xfnm[#1]{\unskip,\space#1}\fi
\bibitem[{sol(2016)}]{soll2016long}
\bibinfo{title}{The long and brutal history of fake news},
  \bibinfo{howpublished}{\url{https://www.politico.com/magazine/story/2016/12/fake-news-history-long-violent-214535}},
  \bibinfo{year}{2016}.
\bibitem[{Wardle and Derakhshan(2017)}]{wardle2017information}
\bibinfo{author}{C.~Wardle}, \bibinfo{author}{H.~Derakhshan},
\newblock \bibinfo{title}{Information disorder: Toward an interdisciplinary
  framework for research and policy making},
\newblock \bibinfo{journal}{Council of Europe report} \bibinfo{volume}{27}
  (\bibinfo{year}{2017}).
\bibitem[{Activities(2017)}]{activities2017intentions}
\bibinfo{author}{A.~R. Activities},
\newblock \bibinfo{title}{Intentions in recent us elections},
\newblock \bibinfo{journal}{Intelligence Community Assessment, Office of the
  Director of National Intelligence} \bibinfo{volume}{6}
  (\bibinfo{year}{2017}).
\bibitem[{Howell et~al.(2013)}]{howell2013digital}
\bibinfo{author}{L.~Howell}, et~al.,
\newblock \bibinfo{title}{Digital wildfires in a hyperconnected world},
\newblock \bibinfo{journal}{WEF report} \bibinfo{volume}{3}
  (\bibinfo{year}{2013}) \bibinfo{pages}{15--94}.
\bibitem[{Barstow(2008)}]{barstow2008behind}
\bibinfo{author}{D.~Barstow},
\newblock \bibinfo{title}{Behind tv analysts, pentagon’s hidden hand},
\newblock \bibinfo{journal}{New York Times} \bibinfo{volume}{20}
  (\bibinfo{year}{2008}) \bibinfo{pages}{A1}.
\bibitem[{Allcott and Gentzkow(2017)}]{allcott2017social}
\bibinfo{author}{H.~Allcott}, \bibinfo{author}{M.~Gentzkow},
\newblock \bibinfo{title}{Social media and fake news in the 2016 election},
\newblock \bibinfo{journal}{Journal of economic perspectives}
  \bibinfo{volume}{31} (\bibinfo{year}{2017}) \bibinfo{pages}{211--36}.
\bibitem[{Quandt et~al.(2019)Quandt, Frischlich, Boberg, and
  Schatto-Eckrodt}]{quandt2019fake}
\bibinfo{author}{T.~Quandt}, \bibinfo{author}{L.~Frischlich},
  \bibinfo{author}{S.~Boberg}, \bibinfo{author}{T.~Schatto-Eckrodt},
\newblock \bibinfo{title}{Fake news},
\newblock \bibinfo{journal}{The international encyclopedia of Journalism
  Studies}  (\bibinfo{year}{2019}) \bibinfo{pages}{1--6}.
\bibitem[{Waldrop(2017)}]{waldrop2017news}
\bibinfo{author}{M.~M. Waldrop},
\newblock \bibinfo{title}{News feature: The genuine problem of fake news},
\newblock \bibinfo{journal}{Proceedings of the National Academy of Sciences}
  \bibinfo{volume}{114} (\bibinfo{year}{2017}) \bibinfo{pages}{12631--12634}.
\bibitem[{Kshetri and Voas(2017)}]{kshetri2017economics}
\bibinfo{author}{N.~Kshetri}, \bibinfo{author}{J.~Voas},
\newblock \bibinfo{title}{The economics of “fake news”},
\newblock \bibinfo{journal}{IT Professional} \bibinfo{volume}{19}
  (\bibinfo{year}{2017}) \bibinfo{pages}{8--12}.
\bibitem[{Braun and Eklund(2019)}]{braun2019fake}
\bibinfo{author}{J.~A. Braun}, \bibinfo{author}{J.~L. Eklund},
\newblock \bibinfo{title}{Fake news, real money: Ad tech platforms,
  profit-driven hoaxes, and the business of journalism},
\newblock \bibinfo{journal}{Digital Journalism} \bibinfo{volume}{7}
  (\bibinfo{year}{2019}) \bibinfo{pages}{1--21}.
\bibitem[{Wingfield et~al.(2016)Wingfield, Isaac, and
  Benner}]{wingfield2016google}
\bibinfo{author}{N.~Wingfield}, \bibinfo{author}{M.~Isaac},
  \bibinfo{author}{K.~Benner},
\newblock \bibinfo{title}{Google and facebook take aim at fake news sites},
\newblock \bibinfo{journal}{The New York Times} \bibinfo{volume}{11}
  (\bibinfo{year}{2016}) \bibinfo{pages}{12}.
\bibitem[{Koohikamali and Sidorova(2017)}]{koohikamali2017information}
\bibinfo{author}{M.~Koohikamali}, \bibinfo{author}{A.~Sidorova},
\newblock \bibinfo{title}{Information re-sharing on social network sites in the
  age of fake news.},
\newblock \bibinfo{journal}{Informing Science} \bibinfo{volume}{20}
  (\bibinfo{year}{2017}).
\bibitem[{Bovet and Makse(2019)}]{bovet2019influence}
\bibinfo{author}{A.~Bovet}, \bibinfo{author}{H.~A. Makse},
\newblock \bibinfo{title}{Influence of fake news in twitter during the 2016 us
  presidential election},
\newblock \bibinfo{journal}{Nature communications} \bibinfo{volume}{10}
  (\bibinfo{year}{2019}) \bibinfo{pages}{1--14}.
\bibitem[{Shu et~al.(2019)Shu, Bernard, and Liu}]{shu2019studying}
\bibinfo{author}{K.~Shu}, \bibinfo{author}{H.~R. Bernard},
  \bibinfo{author}{H.~Liu},
\newblock \bibinfo{title}{Studying fake news via network analysis: detection
  and mitigation},
\newblock in: \bibinfo{booktitle}{Emerging Research Challenges and
  Opportunities in Computational Social Network Analysis and Mining},
  \bibinfo{publisher}{Springer}, \bibinfo{year}{2019}, pp.
  \bibinfo{pages}{43--65}.
\bibitem[{Lazer et~al.(2018)Lazer, Baum, Benkler, Berinsky, Greenhill, Menczer,
  Metzger, Nyhan, Pennycook, Rothschild et~al.}]{lazer2018science}
\bibinfo{author}{D.~M. Lazer}, \bibinfo{author}{M.~A. Baum},
  \bibinfo{author}{Y.~Benkler}, \bibinfo{author}{A.~J. Berinsky},
  \bibinfo{author}{K.~M. Greenhill}, \bibinfo{author}{F.~Menczer},
  \bibinfo{author}{M.~J. Metzger}, \bibinfo{author}{B.~Nyhan},
  \bibinfo{author}{G.~Pennycook}, \bibinfo{author}{D.~Rothschild}, et~al.,
\newblock \bibinfo{title}{The science of fake news},
\newblock \bibinfo{journal}{Science} \bibinfo{volume}{359}
  (\bibinfo{year}{2018}) \bibinfo{pages}{1094--1096}.
\bibitem[{Zhou and Zafarani(2020)}]{zhou2020survey}
\bibinfo{author}{X.~Zhou}, \bibinfo{author}{R.~Zafarani},
\newblock \bibinfo{title}{A survey of fake news: Fundamental theories,
  detection methods, and opportunities},
\newblock \bibinfo{journal}{ACM Computing Surveys (CSUR)} \bibinfo{volume}{53}
  (\bibinfo{year}{2020}) \bibinfo{pages}{1--40}.
\bibitem[{Castillo et~al.(2011)Castillo, Mendoza, and
  Poblete}]{castillo2011information}
\bibinfo{author}{C.~Castillo}, \bibinfo{author}{M.~Mendoza},
  \bibinfo{author}{B.~Poblete},
\newblock \bibinfo{title}{Information credibility on twitter},
\newblock in: \bibinfo{booktitle}{Proceedings of the 20th international
  conference on World wide web}, \bibinfo{year}{2011}, pp.
  \bibinfo{pages}{675--684}.
\bibitem[{Jin et~al.(2016)Jin, Cao, Zhang, and Luo}]{jin2016news}
\bibinfo{author}{Z.~Jin}, \bibinfo{author}{J.~Cao}, \bibinfo{author}{Y.~Zhang},
  \bibinfo{author}{J.~Luo},
\newblock \bibinfo{title}{News verification by exploiting conflicting social
  viewpoints in microblogs},
\newblock in: \bibinfo{booktitle}{Proceedings of the AAAI Conference on
  Artificial Intelligence}, volume~\bibinfo{volume}{30}, \bibinfo{year}{2016}.
\bibitem[{P{\'e}rez-Rosas et~al.(2018)P{\'e}rez-Rosas, Kleinberg, Lefevre, and
  Mihalcea}]{umich}
\bibinfo{author}{V.~P{\'e}rez-Rosas}, \bibinfo{author}{B.~Kleinberg},
  \bibinfo{author}{A.~Lefevre}, \bibinfo{author}{R.~Mihalcea},
\newblock \bibinfo{title}{Automatic detection of fake news},
\newblock in: \bibinfo{booktitle}{Proceedings of the 27th International
  Conference on Computational Linguistics}, \bibinfo{publisher}{Association for
  Computational Linguistics}, \bibinfo{year}{2018}.
\bibitem[{new(2020)}]{newsguard}
\bibinfo{title}{Newsguard: The internet trust tool},
  \bibinfo{howpublished}{\url{https://www.newsguardtech.com}},
  \bibinfo{year}{Accessed: 20.12.2020}.
\bibitem[{Baly et~al.(2018)Baly, Karadzhov, Alexandrov, Glass, and
  Nakov}]{baly2018predicting}
\bibinfo{author}{R.~Baly}, \bibinfo{author}{G.~Karadzhov},
  \bibinfo{author}{D.~Alexandrov}, \bibinfo{author}{J.~Glass},
  \bibinfo{author}{P.~Nakov},
\newblock \bibinfo{title}{Predicting factuality of reporting and bias of news
  media sources},
\newblock in: \bibinfo{booktitle}{Proceedings of the 2018 Conference on
  Empirical Methods in Natural Language Processing},
  \bibinfo{publisher}{Association for Computational Linguistics},
  \bibinfo{address}{Brussels, Belgium}, \bibinfo{year}{2018}, pp.
  \bibinfo{pages}{3528--3539}. \URLprefix
  \url{https://www.aclweb.org/anthology/D18-1389}.
  \DOIprefix\doi{10.18653/v1/D18-1389}.
\bibitem[{Botnevik et~al.(2020)Botnevik, Sakariassen, and
  Setty}]{botnevik2020brenda}
\bibinfo{author}{B.~Botnevik}, \bibinfo{author}{E.~Sakariassen},
  \bibinfo{author}{V.~Setty},
\newblock \bibinfo{title}{Brenda: Browser extension for fake news detection},
\newblock in: \bibinfo{booktitle}{Proceedings of the 43rd International ACM
  SIGIR Conference on Research and Development in Information Retrieval}, SIGIR
  '20, \bibinfo{publisher}{Association for Computing Machinery},
  \bibinfo{address}{New York, NY, USA}, \bibinfo{year}{2020}, p.
  \bibinfo{pages}{2117–2120}.
\bibitem[{Page et~al.(1999)Page, Brin, Motwani, and
  Winograd}]{page1999pagerank}
\bibinfo{author}{L.~Page}, \bibinfo{author}{S.~Brin},
  \bibinfo{author}{R.~Motwani}, \bibinfo{author}{T.~Winograd},
  \bibinfo{title}{The PageRank citation ranking: Bringing order to the web.},
  \bibinfo{type}{Technical Report}, Stanford InfoLab, \bibinfo{year}{1999}.
\bibitem[{med(2021)}]{mediabiasfactcheck}
\bibinfo{title}{Mediabias\/factcheck},
  \bibinfo{howpublished}{\url{https://mediabiasfactcheck.com/}},
  \bibinfo{year}{Accessed: 26.02.2021}.
\bibitem[{Esteves et~al.(2018)Esteves, Reddy, Chawla, and Lehmann}]{jens}
\bibinfo{author}{D.~Esteves}, \bibinfo{author}{A.~J. Reddy},
  \bibinfo{author}{P.~Chawla}, \bibinfo{author}{J.~Lehmann},
\newblock \bibinfo{title}{Belittling the source: Trustworthiness indicators to
  obfuscate fake news on the web},
\newblock in: \bibinfo{booktitle}{Proceedings of the First Workshop on Fact
  Extraction and {VER}ification ({FEVER})}, \bibinfo{publisher}{Association for
  Computational Linguistics}, \bibinfo{year}{2018}.
\bibitem[{Rashkin et~al.(2017)Rashkin, Choi, Jang, Volkova, and Choi}]{rashkin}
\bibinfo{author}{H.~Rashkin}, \bibinfo{author}{E.~Choi}, \bibinfo{author}{J.~Y.
  Jang}, \bibinfo{author}{S.~Volkova}, \bibinfo{author}{Y.~Choi},
\newblock \bibinfo{title}{Truth of varying shades: Analyzing language in fake
  news and political fact-checking},
\newblock in: \bibinfo{booktitle}{Proceedings of the 2017 Conference on
  Empirical Methods in Natural Language Processing},
  \bibinfo{publisher}{Association for Computational Linguistics},
  \bibinfo{year}{2017}.
\bibitem[{Wang(2017)}]{wang}
\bibinfo{author}{W.~Y. Wang},
\newblock \bibinfo{title}{{``}liar, liar pants on fire{''}: A new benchmark
  dataset for fake news detection},
\newblock in: \bibinfo{booktitle}{Proceedings of the 55th Annual Meeting of the
  Association for Computational Linguistics (Volume 2: Short Papers)},
  \bibinfo{publisher}{Association for Computational Linguistics},
  \bibinfo{year}{2017}.
\bibitem[{cor(2020)}]{correctiv}
\bibinfo{title}{Correctiv: Investigations in the public interest},
  \bibinfo{howpublished}{\url{https://correctiv.org}}, \bibinfo{year}{Accessed:
  20.12.2020}.
\bibitem[{Institute(2020)}]{politifact}
\bibinfo{author}{T.~P. Institute}, \bibinfo{title}{Politifact},
  \bibinfo{howpublished}{\url{https://www.politifact.com}},
  \bibinfo{year}{Accessed: 20.12.2020}.
\bibitem[{sno(2020)}]{snopes}
\bibinfo{title}{Snopes}, \bibinfo{howpublished}{\url{https://www.snopes.com}},
  \bibinfo{year}{Accessed: 20.12.2020}.
\bibitem[{Thorne et~al.(2018)Thorne, Vlachos, Cocarascu, Christodoulopoulos,
  and Mittal}]{fever}
\bibinfo{author}{J.~Thorne}, \bibinfo{author}{A.~Vlachos},
  \bibinfo{author}{O.~Cocarascu}, \bibinfo{author}{C.~Christodoulopoulos},
  \bibinfo{author}{A.~Mittal},
\newblock \bibinfo{title}{The fact extraction and {VER}ification ({FEVER})
  shared task},
\newblock in: \bibinfo{booktitle}{Proceedings of the First Workshop on Fact
  Extraction and {VER}ification ({FEVER})}, \bibinfo{publisher}{Association for
  Computational Linguistics}, \bibinfo{year}{2018}, pp. \bibinfo{pages}{1--9}.
\bibitem[{Hossain et~al.(2020)Hossain, Logan~IV, Ugarte, Matsubara, Young, and
  Singh}]{hossain2020covid}
\bibinfo{author}{T.~Hossain}, \bibinfo{author}{R.~L. Logan~IV},
  \bibinfo{author}{A.~Ugarte}, \bibinfo{author}{Y.~Matsubara},
  \bibinfo{author}{S.~Young}, \bibinfo{author}{S.~Singh},
\newblock \bibinfo{title}{{COVIDL}ies: Detecting {COVID}-19 misinformation on
  social media},
\newblock in: \bibinfo{booktitle}{Proceedings of the 1st Workshop on {NLP} for
  {COVID}-19 (Part 2) at {EMNLP} 2020}, \bibinfo{publisher}{Association for
  Computational Linguistics}, \bibinfo{address}{Online}, \bibinfo{year}{2020}.
\bibitem[{Mikolov et~al.(2013)Mikolov, Chen, Corrado, and
  Dean}]{mikolov2013efficient}
\bibinfo{author}{T.~Mikolov}, \bibinfo{author}{K.~Chen},
  \bibinfo{author}{G.~Corrado}, \bibinfo{author}{J.~Dean},
\newblock \bibinfo{title}{Efficient estimation of word representations in
  vector space},
\newblock in: \bibinfo{editor}{Y.~Bengio}, \bibinfo{editor}{Y.~LeCun} (Eds.),
  \bibinfo{booktitle}{1st International Conference on Learning Representations,
  {ICLR} 2013, Scottsdale, Arizona, USA, May 2-4, 2013, Workshop Track
  Proceedings}, \bibinfo{year}{2013}.
\bibitem[{Harris(1954)}]{harris54}
\bibinfo{author}{Z.~Harris},
\newblock \bibinfo{title}{Distributional structure},
\newblock \bibinfo{journal}{Word} \bibinfo{volume}{10} (\bibinfo{year}{1954})
  \bibinfo{pages}{146--162}. \URLprefix
  \url{https://link.springer.com/chapter/10.1007/978-94-009-8467-7_1}.
  \DOIprefix\doi{10.1007/978-94-009-8467-7_1}.
\bibitem[{Firth(1957)}]{Firth_Papers57}
\bibinfo{author}{J.~R. Firth}, \bibinfo{title}{{P}apers in {L}inguistics,
  1934-1951}, \bibinfo{publisher}{{O}xford {U}niversity {P}ress},
  \bibinfo{address}{London}, \bibinfo{year}{1957}.
\bibitem[{Bojanowski et~al.(2017)Bojanowski, Grave, Joulin, and
  Mikolov}]{bojanowski2016enriching}
\bibinfo{author}{P.~Bojanowski}, \bibinfo{author}{E.~Grave},
  \bibinfo{author}{A.~Joulin}, \bibinfo{author}{T.~Mikolov},
\newblock \bibinfo{title}{Enriching word vectors with subword information},
\newblock \bibinfo{journal}{Transactions of the Association for Computational
  Linguistics}  (\bibinfo{year}{2017}) \bibinfo{pages}{135--146}.
\bibitem[{Vaswani et~al.(2017)Vaswani, Shazeer, Parmar, Uszkoreit, Jones,
  Gomez, Kaiser, and Polosukhin}]{vaswani2017attention}
\bibinfo{author}{A.~Vaswani}, \bibinfo{author}{N.~Shazeer},
  \bibinfo{author}{N.~Parmar}, \bibinfo{author}{J.~Uszkoreit},
  \bibinfo{author}{L.~Jones}, \bibinfo{author}{A.~N. Gomez},
  \bibinfo{author}{{\L}.~Kaiser}, \bibinfo{author}{I.~Polosukhin},
\newblock \bibinfo{title}{Attention is all you need},
\newblock in: \bibinfo{booktitle}{Advances in neural information processing
  systems}, \bibinfo{year}{2017}, pp. \bibinfo{pages}{5998--6008}.
\bibitem[{Li and Yang(2018)}]{li2018word}
\bibinfo{author}{Y.~Li}, \bibinfo{author}{T.~Yang},
\newblock \bibinfo{title}{Word embedding for understanding natural language: a
  survey},
\newblock in: \bibinfo{booktitle}{Guide to Big Data Applications},
  \bibinfo{publisher}{Springer}, \bibinfo{year}{2018}, pp.
  \bibinfo{pages}{83--104}.
\bibitem[{Almeida and Xex{\'e}o(2019)}]{almeida2019word}
\bibinfo{author}{F.~Almeida}, \bibinfo{author}{G.~Xex{\'e}o},
\newblock \bibinfo{title}{Word embeddings: A survey},
\newblock \bibinfo{journal}{arXiv preprint arXiv:1901.09069}
  (\bibinfo{year}{2019}).
\bibitem[{Kusner et~al.(2015)Kusner, Sun, Kolkin, and
  Weinberger}]{kusner2015word}
\bibinfo{author}{M.~Kusner}, \bibinfo{author}{Y.~Sun},
  \bibinfo{author}{N.~Kolkin}, \bibinfo{author}{K.~Weinberger},
\newblock \bibinfo{title}{From word embeddings to document distances},
\newblock in: \bibinfo{booktitle}{International conference on machine
  learning}, \bibinfo{year}{2015}, pp. \bibinfo{pages}{957--966}.
\bibitem[{Balikas et~al.(2018)Balikas, Laclau, Redko, and Amini}]{regWMD}
\bibinfo{author}{G.~Balikas}, \bibinfo{author}{C.~Laclau},
  \bibinfo{author}{I.~Redko}, \bibinfo{author}{M.-R. Amini},
\newblock \bibinfo{title}{Cross-lingual document retrieval using regularized
  wasserstein distance},
\newblock in: \bibinfo{booktitle}{Proceedings of the 40th European Conference
  {ECIR} conference on Information Retrieval, {ECIR} 2018, Grenoble, France,
  March 26-29, 2018}, \bibinfo{year}{2018}.
\bibitem[{Shu et~al.(2020)Shu, Mahudeswaran, Wang, Lee, and Liu}]{data}
\bibinfo{author}{K.~Shu}, \bibinfo{author}{D.~Mahudeswaran},
  \bibinfo{author}{S.~Wang}, \bibinfo{author}{D.~Lee},
  \bibinfo{author}{H.~Liu},
\newblock \bibinfo{title}{Fakenewsnet: A data repository with news content,
  social context, and spatiotemporal information for studying fake news on
  social media},
\newblock \bibinfo{journal}{Big Data}  (\bibinfo{year}{2020}).
\bibitem[{Yokoi et~al.(2020)Yokoi, Takahashi, Akama, Suzuki, and
  Inui}]{yokoi2020word}
\bibinfo{author}{S.~Yokoi}, \bibinfo{author}{R.~Takahashi},
  \bibinfo{author}{R.~Akama}, \bibinfo{author}{J.~Suzuki},
  \bibinfo{author}{K.~Inui},
\newblock \bibinfo{title}{Word rotator’s distance},
\newblock in: \bibinfo{booktitle}{Proceedings of the 2020 Conference on
  Empirical Methods in Natural Language Processing (EMNLP)},
  \bibinfo{year}{2020}, pp. \bibinfo{pages}{2944--2960}.

\end{thebibliography}

\end{document}